\documentclass[10pt, conference, compsocconf]{IEEEtran}

\ifCLASSINFOpdf
\else
\fi

\usepackage{cite}
\usepackage{amsmath,amssymb,amsfonts}
\usepackage{algorithmic}
\usepackage{graphicx}
\usepackage{textcomp}
\usepackage{xcolor}
\usepackage{caption}
\usepackage{subcaption}
\usepackage{soul}

\newcommand{\dataTrain}{\mathcal{D}_\text{TRAIN}}

\begin{document}
%
\title{Discovering outliers in the Mars Express thermal power consumption patterns}


\author{
\IEEEauthorblockN{Matej Petkovi\'{c}$^{1,2}$, Luke Lucas$^3$, Toma\v{z} Stepi\v{s}nik$^{1,2}$, Pan\v{c}e Panov$^{1,2}$, Nikola Simidjievski$^{1,2,4}$, Dragi Kocev$^{1,2}$}
\IEEEauthorblockA{
\textit{$^1$Bias Variance Labs, Ljubljana, Slovenia} \\
\textit{$^2$Jo\v{z}ef Stefan Institute, Ljubljana, Slovenia} \\
\textit{$^3$LSE Space GmbH, Gilching, Germany} \\
\textit{$^4$University of Cambridge, Cambridge, UK}\\
\{matej.petkovic, tomaz.stepisnik, pance.panov, dragi.kocev\}@ijs.si, luke.lucas@esa.int, ns779@cam.ac.uk}
}


%


\maketitle

\begin{abstract}
The Mars Express (MEX) spacecraft has been orbiting Mars since 2004. The operators need to constantly monitor its behavior and handle sporadic deviations (outliers) from the expected patterns of measurements of quantities that the satellite is sending to Earth. In this paper, we analyze the patterns of the electrical power consumption of  MEX's thermal subsystem, that maintains the spacecraft's temperature at the desired level. The consumption is not constant, but should be roughly periodic in the short term, with the period that corresponds to one orbit around Mars. By using long short-term memory neural networks, we show that the consumption pattern is more irregular than expected, and successfully detect such irregularities, opening possibility for automatic outlier detection on MEX in the future.
\end{abstract}


%
\IEEEpeerreviewmaketitle

\section{Introduction}

Spacecraft' health and endurance depend on close monitoring and accurate analysis of their telemetry data. This is especially important for older but still operational spacecraft, that have been operating in harsh environments for very long time equipped with limited computational capabilities, decaying batteries and components (e.g., processors and memory) lagging decades behind the current state-of-the-art. Analyzing these data is non-trivial. Telemetry data are heterogeneous and complex, comprised from measurements and activity records from the different on-board equipment and sensors, typically noisy and incomplete. In turn, operators need to constantly monitor and analyze them, handling sporadic deviations (outliers) from the expected patterns of measurements that relate to the spacecraft's behavior.

Outlier (or anomaly) detection refers to identification and investigation of rare (and unexpected) events and patterns in the data, which do not conform to the underlying data distribution. In the context of spacecraft operations, typically such outliers are a result of an on-board equipment malfunction or unexpected (and/or novel) environmental effect. In this work, we analyze telemetry data from the Mars Express (MEX) spacecraft to detect anomalies in the electrical power consumption of MEX's thermal subsystem which maintains the spacecraft's temperature at the desired level. 

MEX, a long-lasting mission of the European Space Agency, has been exploring Mars since 2004. It is responsible for a wealth of scientific data comprised of three-dimensional renders of the surface and a complete map of the chemical composition of Mars’s atmosphere that has led to important scientific discoveries, such as the evidence of the presence of water (above and below the surface of the planet). The scientific payload of MEX consists of seven instruments, which together with the on-board equipment have to be kept within their operating temperature ranges (from room temperature for some instruments, to temperatures as low as $-180^oC$ for others). In order to maintain these predefined operating temperatures, the spacecraft is equipped with an autonomous thermal system composed of 33 heater lines that consume a significant amount of the total generated electric power -- leaving a fraction to be used for science operations. Therefore, given the age and the current condition of MEX, monitoring this consumption and identifying unexpected malfunction has a direct consequence on the longevity of the spacecraft and its mission \cite{lucas2017:mars,smc:2017,mex:mi,boumghar:spaceops,giros}.

We propose a machine learning approach for identifying outliers in the MEX's thermal power consumption patterns. The proposed approach combines several state-of-the art unsupervised machine learning methods for anomaly detection to obtain accurate estimates of anomalous behavior. We evaluate the proposed approach on 11.5 years of MEX data showcasing its potential and practical utility with respect to the identified outliers. 

The paper is organized as follows. Section~\ref{sec:related} discusses the related work. Next, Section~\ref{sec:data} describes the MEX telemetry data. Section~\ref{sec:pipeline} presents the proposed pipeline. Section~\ref{sec:experiments} give the design of the experimental evaluation, while Section~\ref{sec:results} discusses the results. Finally, Section~\ref{sec:conclusions} concludes the paper.

\section{Related Work}\label{sec:related}

Anomaly detection is a very active field of research \cite{chandola09,pang2020} focusing on identifying point anomalies (a given single value that differs from the other values), univariate contextual anomalies ( considering values of single variables in a spatial, or temporal context, i.e., outliers in a single time series) and collective anomalies (when consecutive data points are anomalous with respect to the entire signal) \cite{pilastre2020}. A more challenging task refers to multivariate contextual anomaly detection, where a behavior is considered anomalous only with respect to other behavior. 

In the context of analyzing telemetry data, anomaly detection methods typically focus on out-of-limits checking (comparing the values against predefined optimal operating ranges) and analysis of aggregated statistical features (e.g., daily minimum, maximum and mean for a large set of parameters) \cite{heras:spaceops2012,fuertes:spaceops2016,heras2014:jrnl}. The former is used systematically upon the reception of new telemetry data, while the latter is used at a scheduled time intervals (e.g., monthly). 

Other machine learning approaches include k-nearest neighbors (i.e., comparisons of local outlier probabilities) for distinguishing between novel behavior and anomalies in XMM-Newton and Venus Express; and using support vector machines for monitoring the status of a CNES-operated spacecraft. Furthermore, Yairi et al. \cite{Yairi17:jrnl} propose a probabilistic clustering method for detecting measurement anomalies in JAXA's SDS-4 spacecraft, while Carlton et al. \cite{Carlton2018:jrnl} analyze the telemetry data from 32 geostationary Earth Orbit spacecraft using transient event detection and change point event detection methods.

More sophisticated machine learning approaches include generative deep neural networks (Variational Autoencoders and Generative Adversarial Networks) for detecting anomalies in the LUNar Attitude and Orbit Control System (AOCS) SIMulator (LUNASIM) \cite{Ahn2020:jrnl}. Similarly, there are several recent attempts at using long short-termmemory (LSTMs) networks (a type of recurrent neural networks) for the task of anomaly detection. Hundman et al. \cite{hudman2018} investigate the potential of LSTMs for predicting spacecraft telemetry focusing on interpretability, scale, precision, and complexity -- properties inherent in many anomaly detection scenarios. They use expert-labeled data derived from Incident Surprise, Anomaly reports for the Mars Science Laboratory rover, Curiosity, and the Soil Moisture Active Passive satellite. Pan et al. \cite{Pan2020} propose to use bidirectional LSTMs to transform the telemetry data and approach them as a task of supervised regression and time series prediction. The approach is evaluated on data from the SMAP satellite. Finally, Chen et al. \cite{chen2021} propose Bayesian LSTMs that are based on bidirectional LSTMs, where Monte Carlo sampling variance, prediction entropy and mutual information are used to assess the uncertainty of the outputs, and use variational autoencoder to re-evaluate the samples with high uncertainty -- this improves the anomaly detection performance. The approach was evaluation using telemetry data from an anonymous satellite power module system.

\section{Materials and methods}

\subsection{Data}\label{sec:data}

The analyzed data contains values of electrical currents running through the 33 electrical heaters on MEX that are part of the MEX thermal subsystem, spanning approximately 11.5 years, from 2008-08-22 to 2020-01-17. In our analyses we sum the values of the individual heater lines in a total current $x(t)$. The currents are logged regularly at the frequency of approximately two measurements per minute. However, such a fine granularity is not practically useful (the values change too abruptly), therefore we analyze the time series $x(t)$ on the level of $15$-minute intervals (as suggested in \cite{mex:mi}). Each interval $[t_i, t_{i + 1})$ ($t_{i + 1} - t_i = 15 \min$) is assigned a value $x(t_i)$, which is the average value of $x(t)$ for that time interval.

\begin{figure}[!t]
    \centering
    \includegraphics[width=\linewidth]{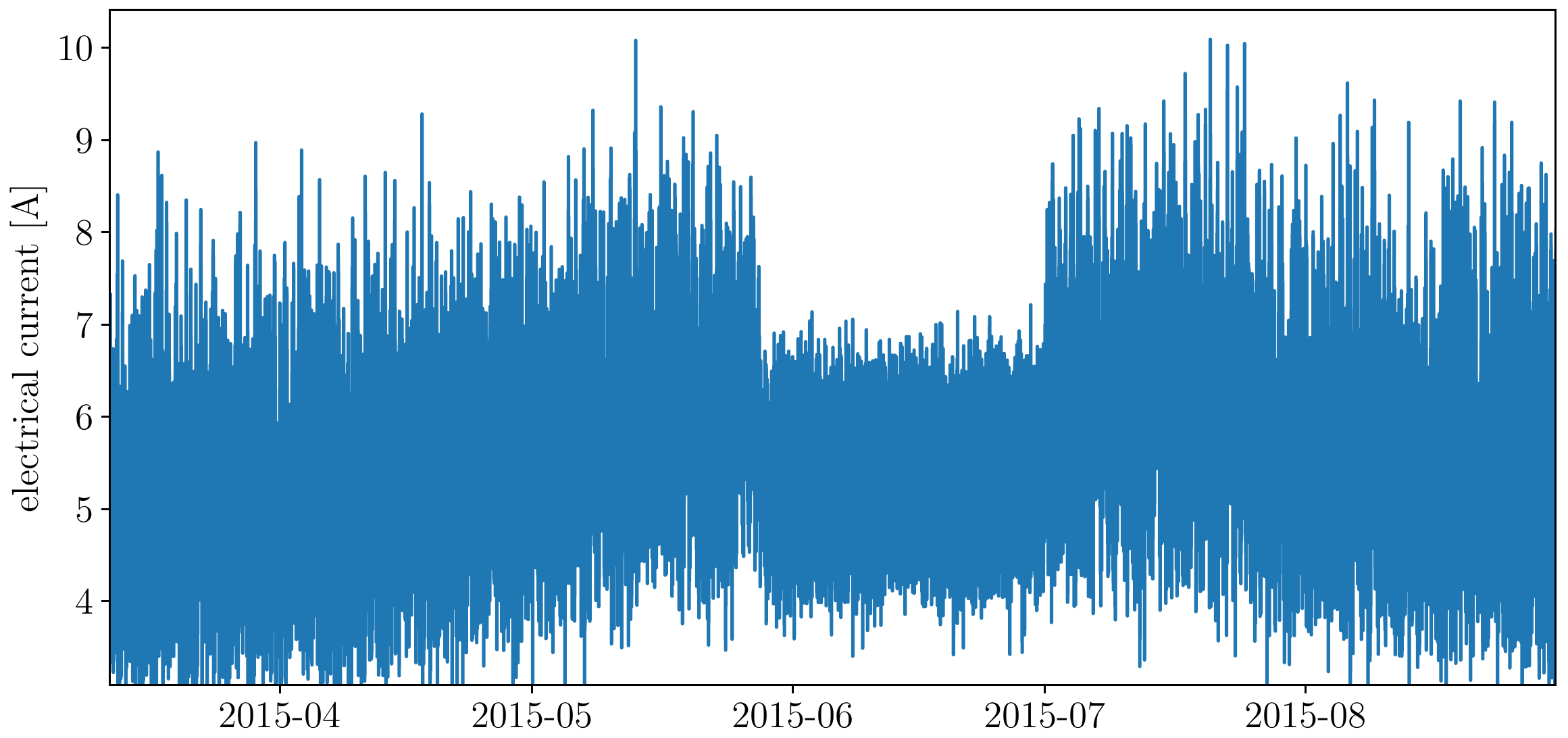}
    \caption{The conjunction of Mars and the Sun in June 2015 was the reason for substantially different behavior of the MEX thermal subsystem.}
    \label{fig:conjunction}
\end{figure}

On short term, we assume that the values of electrical currents should be roughly periodic with the period that corresponds to one MEX orbit around Mars. Therefore, we use one orbit as the unit of analysis. However this assumption is not true for the long term. In the long term, orbital mechanics such as the 26 month Mars year define another period over which the thermal consumption changes. For example, when Sun and Mars are in a conjunction (Sun being between Earth and Mars), the operations on MEX are suspended as communication is disrupted by the Sun. This may considerably change the behavior of the thermal subsystem, as shown in Figure~\ref{fig:conjunction}.

The data describing the orbits (in terms of start and end) are given as part of the events data, as crossings through apocenter and pericenter of its highly elliptical orbit. More specifically, two consecutive crossings through the apocenter define one orbit. Inspection shows that all but two of $16,783$ orbits take approximately $6.75$ hours, i.e., one orbit corresponds to $N = 27$ measurements $x(t_i)$ of the electrical current. Since the methods that we use demand that each example (unit of analysis) has the same number of measurements, we set this number to 27. A unit of analysis is thus a curve
$$x_i = [x(t_i), x(t_{i + 1}), \dots, x(t_{i + N - 1})],$$ and a machine learning method should learn to distinguish abnormal shapes of the curve from normal ones.

\begin{figure}[!t]
    \centering
    \includegraphics[width=\linewidth]{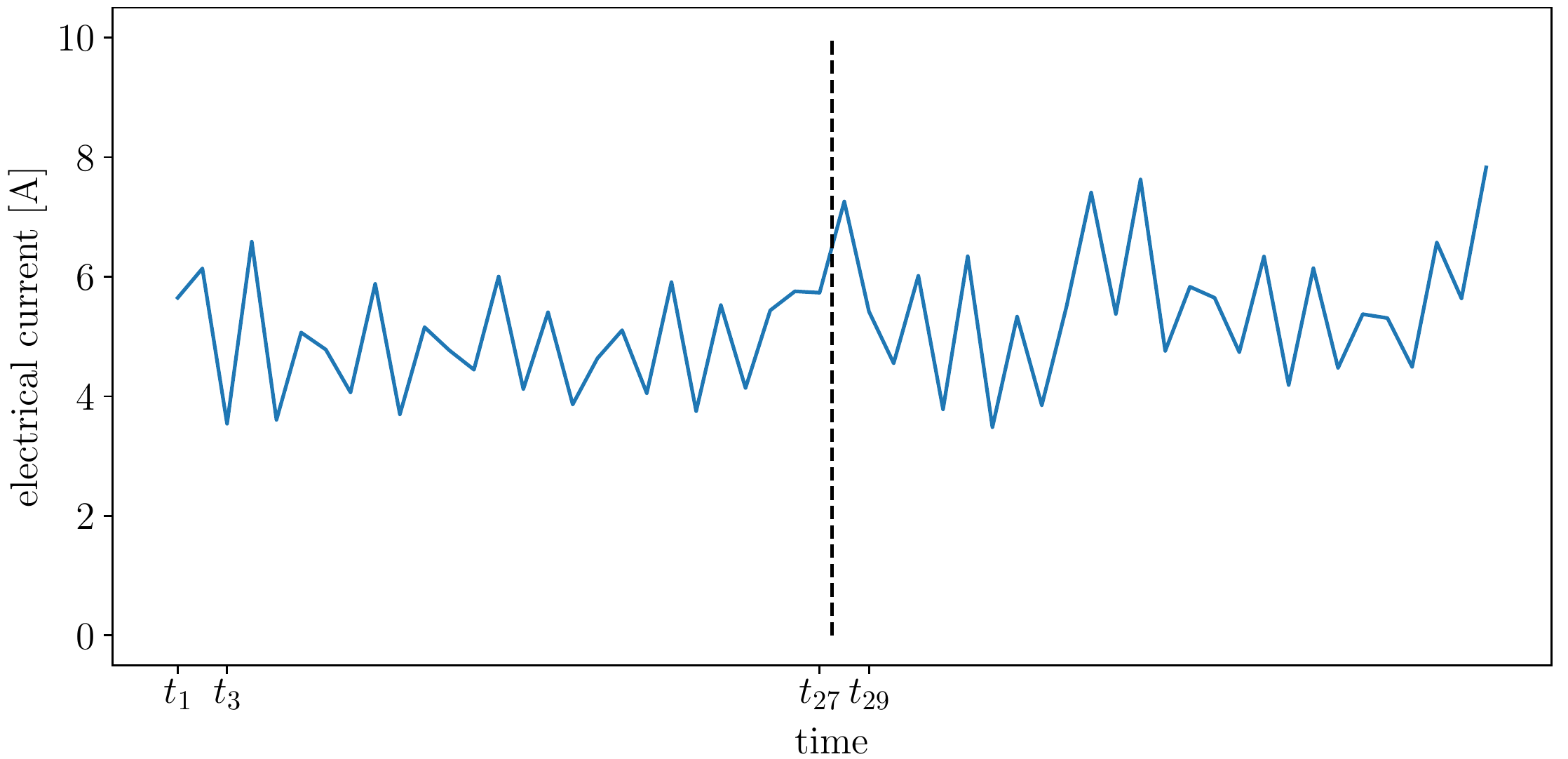}
    \caption{Electrical current through the first two orbits from the data.
    The vertical line lies between the end of the first orbit and the start of the second one. The first data example $x_1$ contains the measurements taken between $t_1$ and $t_{27}$ (inclusive). The third data example $x_2$ contains the measurements taken between $t_3$ and $t_{29}$.
    }
    \label{fig:twoOrbits}
\end{figure}

For example, Fig.~\ref{fig:twoOrbits} shows the first two orbits from the data.
Note that, since one example includes 27 measurements (in an orbit), consecutive examples ($x_i$, $x_{i + 1}$, $x_{i + 2}$ \dots) overlap. For instance, $x_1$ contains the measurements taken between $t_1$ and $t_{27}$, while $x_3$ contains the measurements taken between $t_3$ and $t_{29}$. By introducing overlaps, we explicitly include the temporal information directly into the data which is appropriate for methods such as k-means and isolation forests. The final dataset contains 369,843 examples $x_i$, which roughly equals the number of measurements and not the number of orbits. In the reminder of the paper, $x_i\in \mathbb{R}^N$ denotes an example in the data. 

\subsection{Methods}\label{sec:pipeline}

\begin{figure*}[!t]
    \centering
    \includegraphics[width=.8\linewidth]{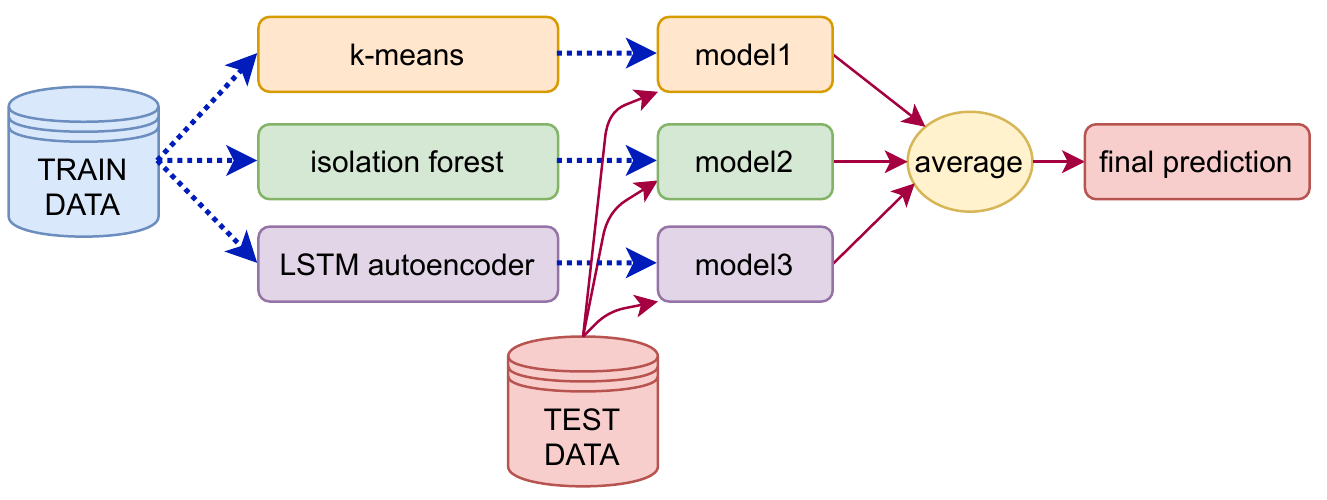}
    \caption{The proposed outlier detection approach: Train data is used to learn three different outlier detection models. During test time, the models are combined into an ensemble, by averaging the individual model prediction into the final, ensemble, prediction.}
    \label{fig:pipeline}
\end{figure*}

We propose constructing a heterogeneous ensemble combining different models for outlier detection. The models are derived from three unsupervised learning algorithms: $k$-means, isolation forest and long short-term memory (LSTM) autoencoders. The complete pipeline is presented in Fig.~\ref{fig:pipeline}. 

\subsubsection{k-means}

$k$-means \cite{kmeans} is an expectation-maximization clustering algorithm. The clustering is performed iteratively: First, the cluster centers $c_j$ (prototypes) $1\leq j \leq k$ are initialized using different procedures such as $k$-means++ initialization \cite{kmeans-pp}, which randomly chooses the centers to be far apart. At each iteration, every data point is assigned its closest prototype, determined by squared Euclidean distance (maximization step). Next, the center of mass of around each prototype (expectation step) becomes the new cluster prototype. The procedure runs until the assignments no longer change, guaranteeing monotonic convergence to a locally optimal point.

A $k$-means outlier score of a given example is defined as
\begin{equation}
    \label{eqn:k-means-score}
  \displaystyle s_\text{k-means}(x_i) = \min_j ||x_i - c_j||_2^2, 
\end{equation}
i.e., the minimal distance between the data point and the final prototypes.

\subsubsection{Isolation Forest}

The motivation for isolation forest \cite{isolation-forest} outlier score is that an example is an outlier if it can be easily isolated from the others by randomly chosen tests. The method constructs many extremely randomized trees, storing (for each example $x_i$ and each tree $t_j$) the path length  $|p_j(x_i)|$ from the root of each tree to the leaf that includes a particular example. In turn, these path lengths are averaged over the trees in the forest resulting in a measure of normality:
\begin{equation}
    \label{eqn:iso-forest-score}
 \displaystyle s_\text{IsoFor}(x_i) = \operatorname{mean}_j |p_j (x_i)|, 
\end{equation}
Higher score (longer average path) corresponds to less anomalous example. 

\subsubsection{LSTM Autoencoder}

\begin{figure*}
    \centering
    \includegraphics[width=1\linewidth]{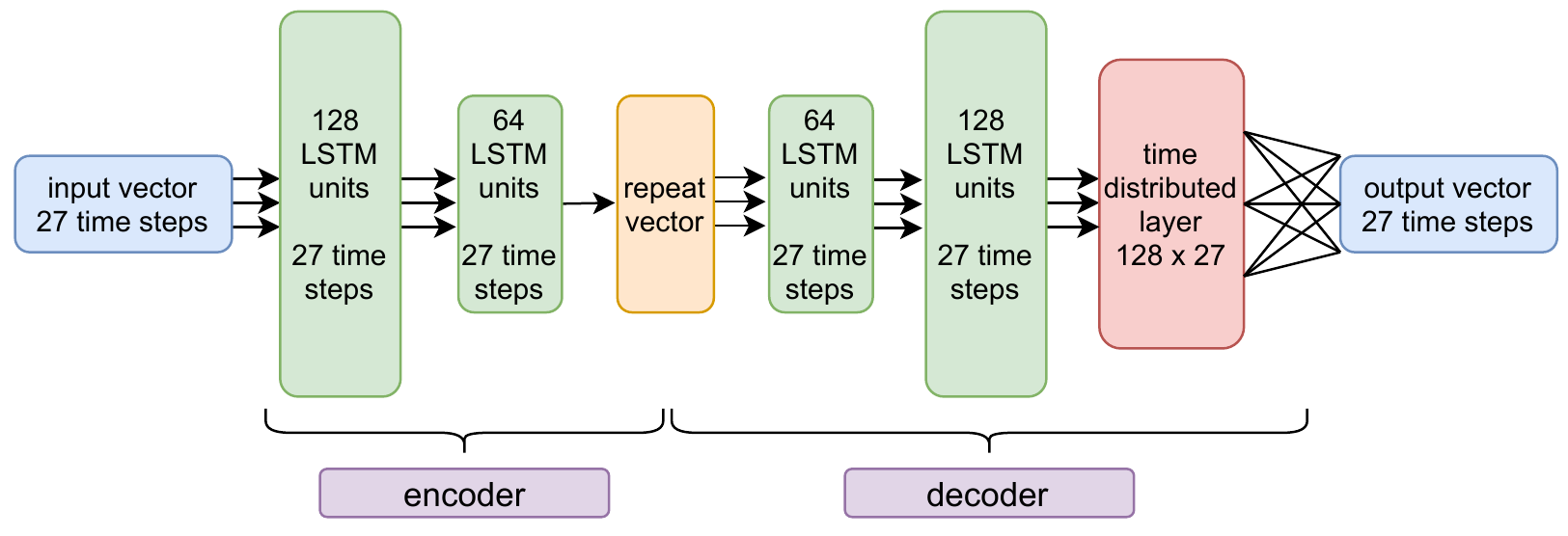}
    \caption{LSTM autoencoder architecture: an input layer, an encoder (comprised of two LSTM layers), a RepeatVector layer, a decoder (with two LSTM layers), time distributed layer and the output layer. Triple arrows denote that a sequence (of length $N = 27$) is passed to the next layer, while a single arrow denotes that only a single number is passed. The RepeatVector layer copies its input $N$ times. The TimeDistributed layer is fully connected to the output layer. The value of $N$ corresponds to the number of data points in one MEX orbit.}
    \label{fig:lstm-ae}
\end{figure*}

While both k-means and isolation forest do not take the time component into account, LSTMs can model temporal data by design \cite{LSTM}. In our approach, we use this property to design LSTM autoencoders for learning latent representations of the data while preserving the long and short-term patterns. As shown in Fig.~\ref{fig:lstm-ae}, every AE consist of two main parts: an encoder and a decoder \cite{autoencoder}. The encoder is a LSTM network that learns a latent code of the data at input. The decoder, on the other hand, takes these codes and attempts decode them. Example which cannot be properly decoded is considered as an outlier. As a measure of abnormality, we use the reconstruction error where larger reconstruction error means more likely that an example is an outlier :

\begin{equation}
    \label{eqn:lstm-score}
    s_\text{LSTM}(x_i) = \operatorname{MSE}(x_i) = ||x_i - \hat{x}_i||_2^2,
\end{equation}
where $\hat{x}_i$ is the output of the network for $x_i$ in input.

\subsection{Ensemble of outlier-detection models}

Ensembles are classical approach for improving the predictive performance by combining predictions from different models. For the task of outlier detection, we combine the individual outlier scores $s(x_i)$ from the three different methods in order to improve the overall performance. Note however, since the outlier scores of different methods are not directly comparable, we first normalize the scores as

\begin{equation*}
    \label{eqn:score-normalization}
     s'_\text{method}(x_i) = 
    \begin{cases}
        \frac{s_\text{method}(x_i) - m_\text{method}}{M_\text{method} - m_\text{method}}\kern-2mm &; \text{method}\neq \text{isolation forest}\\
       \frac{M_\text{method} - s_\text{method}(x_i)}{M_\text{method} - m_\text{method}}\kern-2mm &; \text{method} = \text{isolation forest}
    \end{cases}
\end{equation*}
where $M_\text{method}$ and $m_\text{method}$ are maximal and minimal value of $s_\text{method}$, respectively. This maps the scores to $[0, 1]$ interval, so that $1$ indicates high abnormality.
The final score from the ensemble is then defined as
\begin{equation*}
    \label{eqn:ensemble-score}
   \displaystyle s_\text{ensemble}(x_i) = \frac{1}{3} ( s'_\text{k-means}(x_i) +s'_\text{IsoFor}(x_i) +s'_\text{LSTM}(x_i)\,).
\end{equation*}

\section{Experimental Setup}\label{sec:experiments}

\subsubsection{Parametrisation}

\begin{figure}[!b]
    \centering
    \includegraphics[width=.8\linewidth]{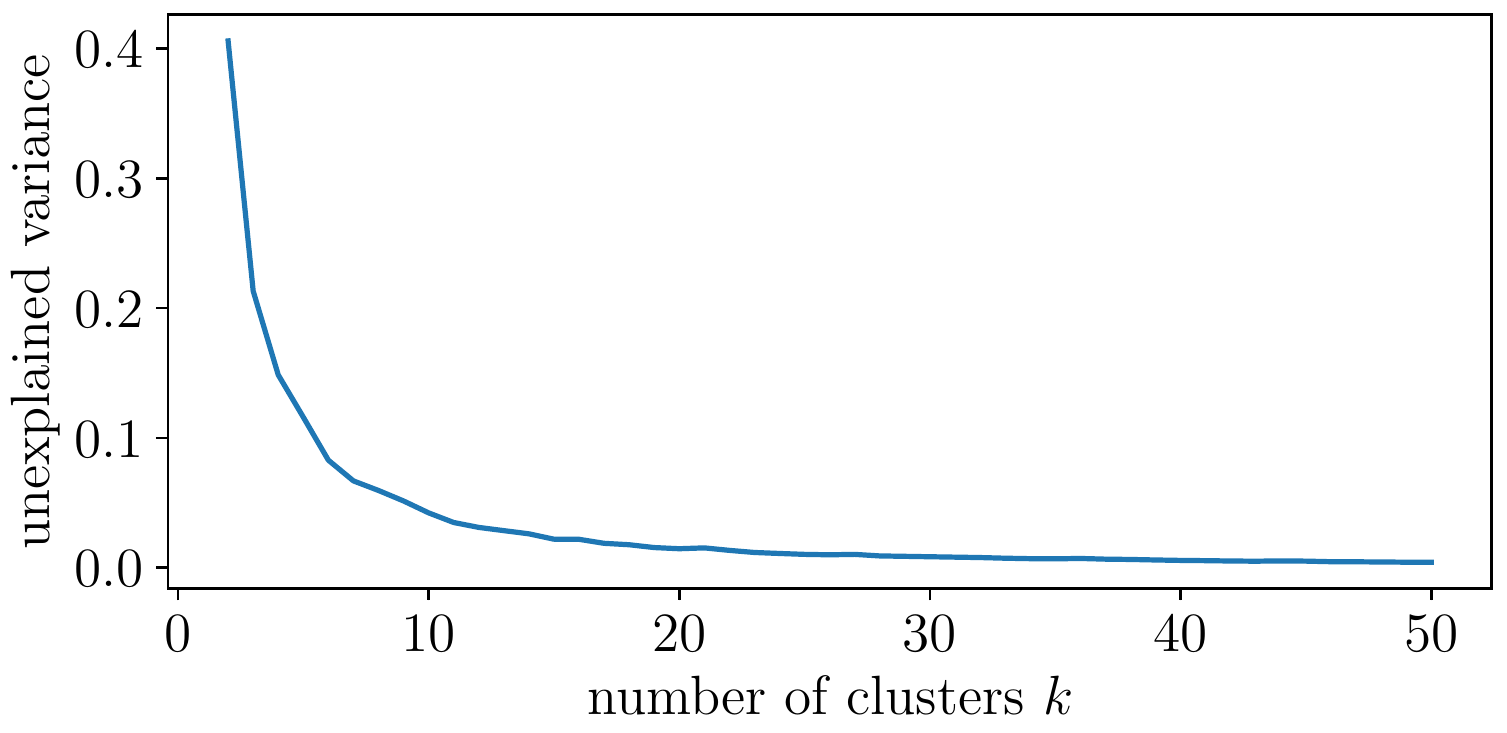}
    \caption{Unexplained variance for different number of k-means clusters.}
    \label{fig:k-means-elbow}
\end{figure}

We set the number of clusters in $k$-means to $k= 50$, which allows for 50 prototypical curves of electrical current. Note that the elbow method for $k$-means (Fig.~\ref{fig:k-means-elbow}) reveals that $k = 30$ prototypical curves could suffice, however $k = 50$ decreases the amount of false positive alarms.

The contamination parameter of Isolation Forests is set to $0.001$, i.e., we expect $0.1\%$ of the examples to be outliers. The remaining parameters are set to the values recommended in \cite{isolation-forest}.

The LSTM autoencoders are implemented using the Keras deep learning library \cite{chollet2015keras}. A more detailed overview of the autoencoder architecture is presented in Figure~\ref{fig:lstm-ae}. It consists of an encoder with two LSTM layers (with 128 and 64 units, respectively) followed by a RepeatVector layer, a decoder with two LSTM layers (with 64 and 128 units) and a fully connected layer at the end. This is a standard autoencoder architecture \cite{autoencoder}. In terms of activation function, we use \textit{ReLU} function, since the preliminary experiments showed better performance than \textit{tanh} activation. The models were trained for $1000$ epochs using the Adam optimizer with a learning rate of $0.001$ (and with the recommended parameters) and a batch size of $128$ (chosen after evaluating batch sizes of $\{2^{5}, 2^6, \dots, 2^{14}\}$). We use the last $20\%$ of examples in the train data for early stopping validation criteria: If no progress has been made in the last 50 epochs, the training stops. The objective function considered is mean-squared error.

\subsubsection{Evaluation procedure}

Since the data spans from 2008 to 2020, we create 12 train sets $\dataTrain{}$ with different lengths. All train sets start on 2008-22-8 but end on December 31st of each year ($y\in\{2009, 2010, \dots, 2020\}$). The respective test sets start where the train set ended and is exactly one (Earth) year long. The exception is the 2020 test set, which is less than 2 months long. For each test set, we also identify the 10 orbits with the highest ensemble score $s_\text{ensemble}$ which are then manually inspected and explained by a MEX spacecraft operator.

\section{Results}\label{sec:results}

\begin{figure}
    \centering
    \includegraphics[width=.93\linewidth]{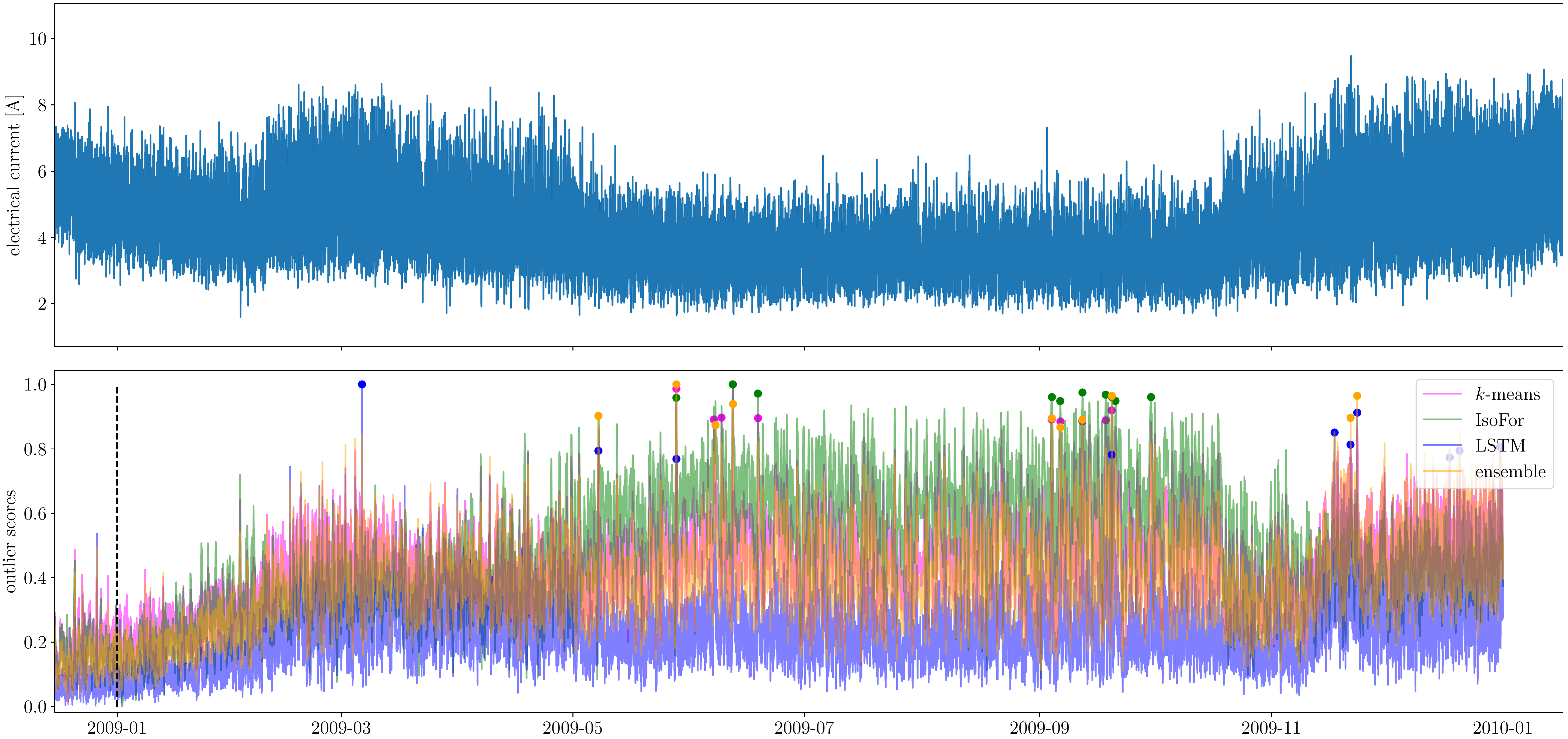}
    \caption{Outliers in the $2009-2010$ period. The border between the train and test data is denoted by the vertical dotted line.}
    \label{fig:2009}
\end{figure}

We start with the test period $2009-2010$, where the first conjunction appeared (note that no conjunction was present in the train data). Figure \ref{fig:2009} shows that the ensemble outlier scores (as well as the single model scores) correctly identify this behavior as anomalous. Additionally, when the electrical current values were unusually high, this was again detected by the ensemble method (but not by some of the single models). Note that the conjunctions in later years are no longer considered anomalous, since the models can learn from the conjunction in 2009.

\begin{figure}
    \centering
    \includegraphics[width=\linewidth]{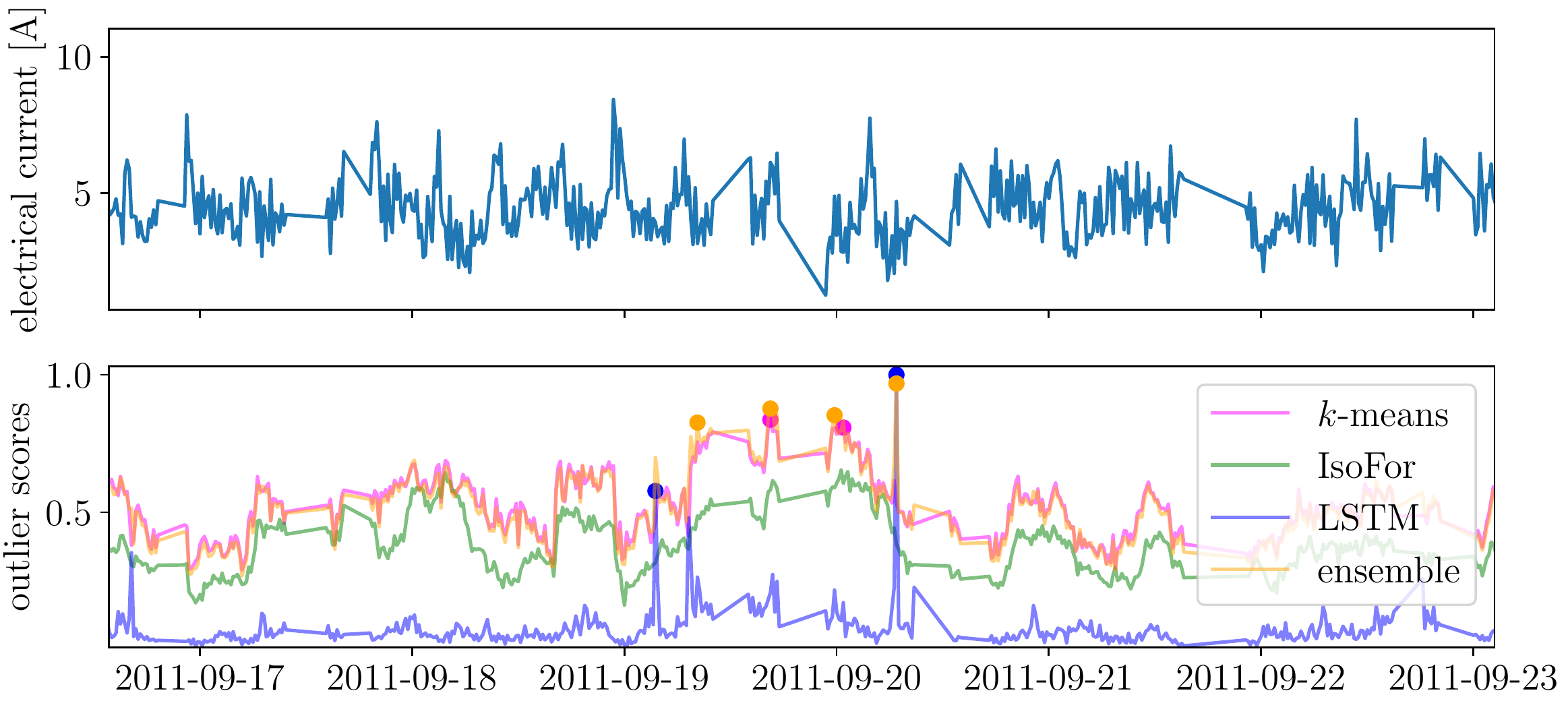}
    \caption{Outliers detected due to missing data in the $2011-2012$ period.}
    \label{fig:2011}
\end{figure}

\begin{figure}
    \centering
    \includegraphics[width=\linewidth]{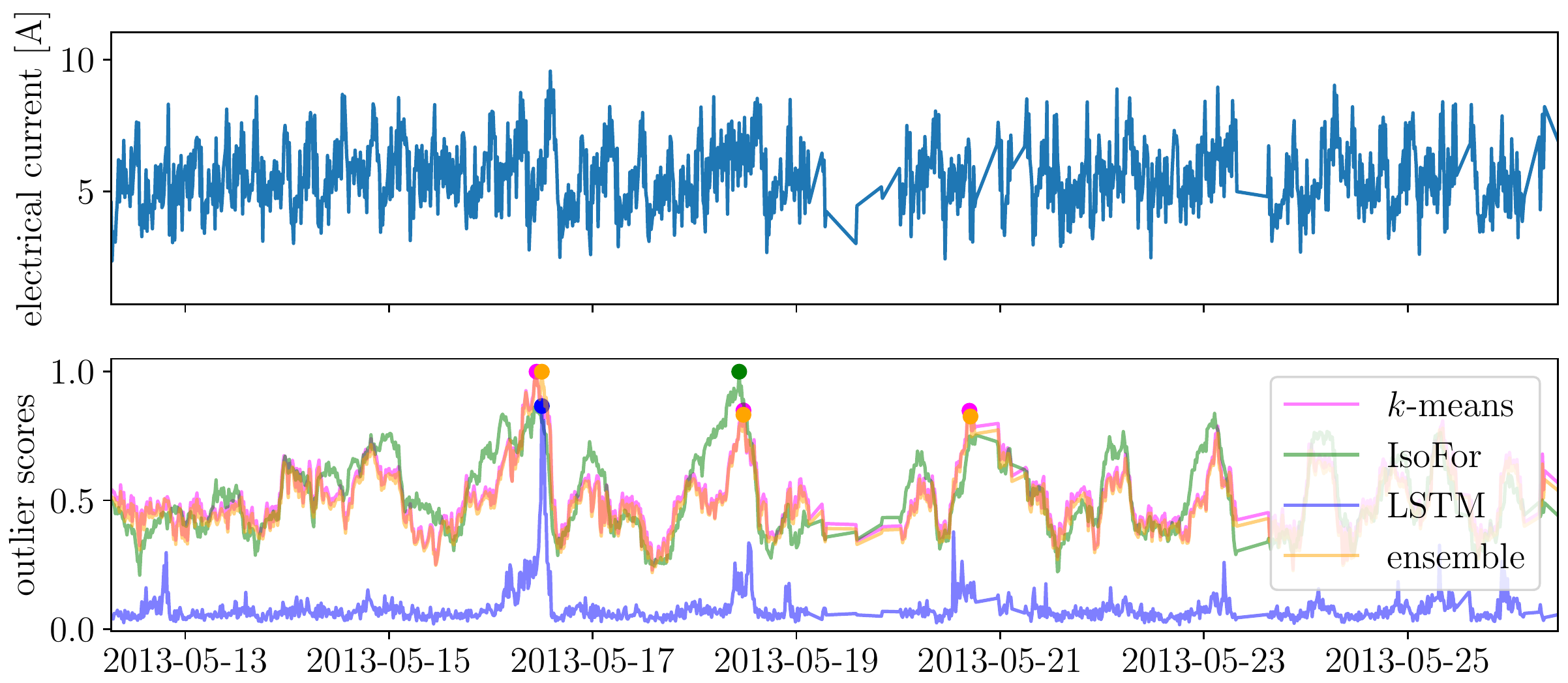}
    \caption{Outliers in May 2013: a) a high peak, followed by quick drop, and b) missing data.}
    \label{fig:2013}
\end{figure}

In the testing period $2011-2012$, most of the top-10 identified outliers are due to the missing values in the data (see Figure~\ref{fig:2011}). Similar behavior can be also observed in the testing period $2013-2014$, however here we also detect phenomenon of an unusually high peak, followed by a quick drop (Fig.~\ref{fig:2013}).

\begin{figure}
    \centering
    \begin{subfigure}[c]{\linewidth}
      \includegraphics[width=\linewidth]{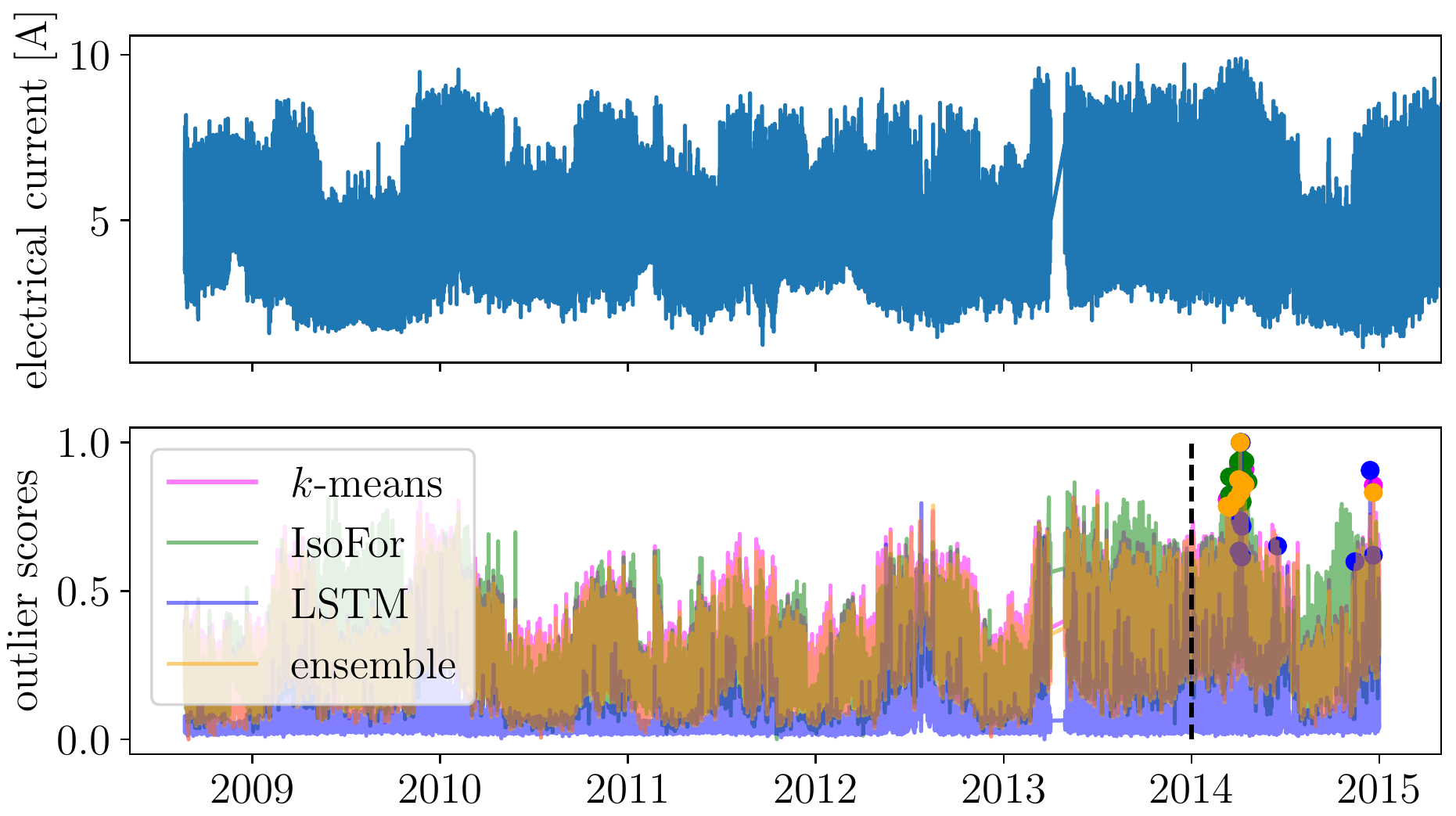}
      \caption{Outliers in $2014$.}\label{fig:2014a}
    \end{subfigure}
    
    \begin{subfigure}[c]{\linewidth}
      \includegraphics[width=\linewidth]{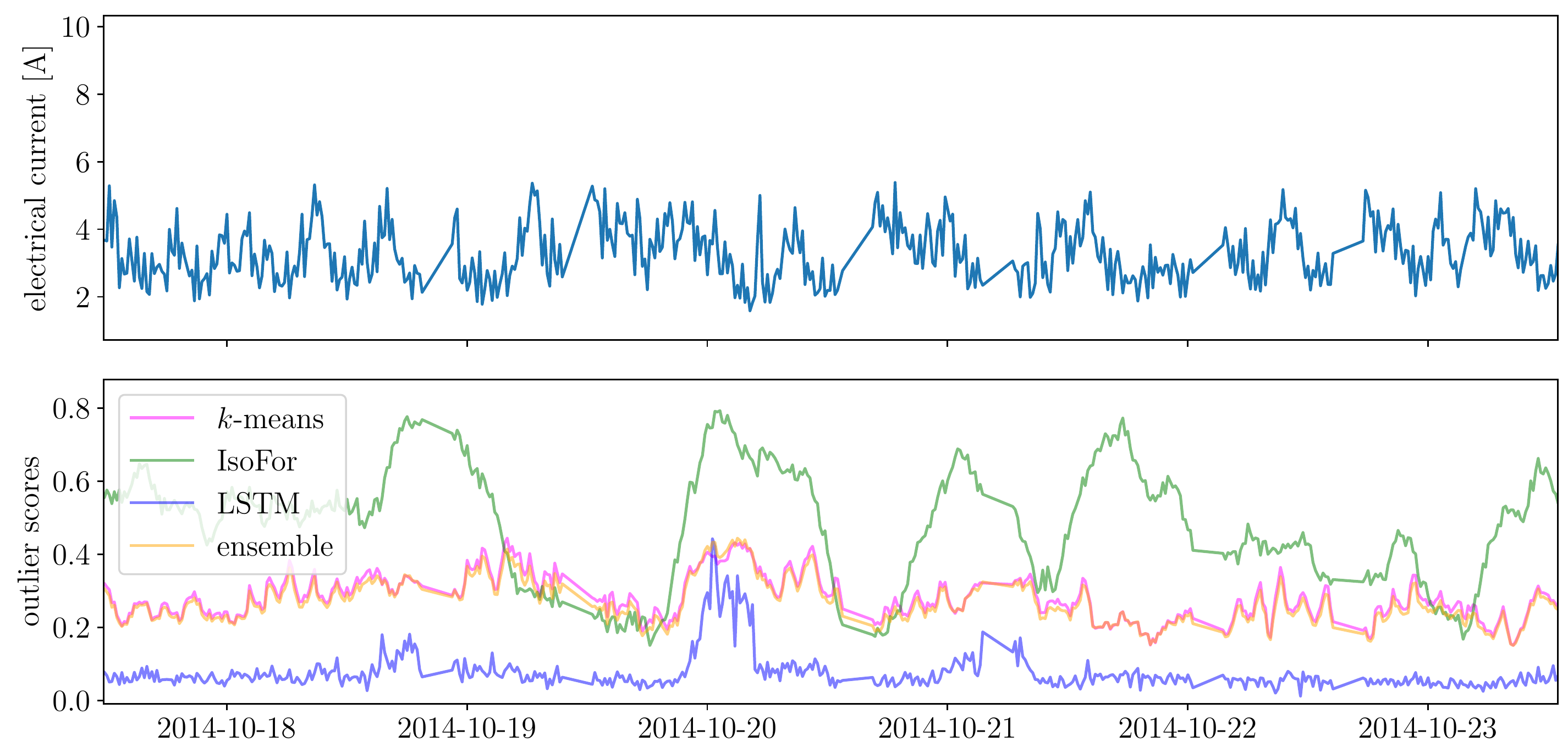}
      \caption{The effect of hiding MEX from the comment Siding Spring }\label{fig:2014b}
    \end{subfigure}
    \caption{(a) Outliers in the consumption in 2014, (b) the effect of hiding MEX from the comment Siding Spring (October 20 2014) was not among the top-10 largest outliers.}
    \label{fig:2014}
\end{figure}

In the testing period $2014-2015$, while the high peaks (highest thus far) of electrical current are successfully detected by our methods (Figure~\ref{fig:2014a}), there might be trade-off of potentially missing the other, less obvious, outliers. To evaluate this, we closely examine the electrical currents during a rare event Figure~\ref{fig:2014b} when MEX undertook special operations to minimise exposure to comet Siding Spring and itsoma and tail. While ESA mission operators executed various novel commands that can lead to increased power consumption, their effects, in fact, are not significant therefore safe and affordable.

\begin{figure}
    \centering
    \includegraphics[width=\linewidth]{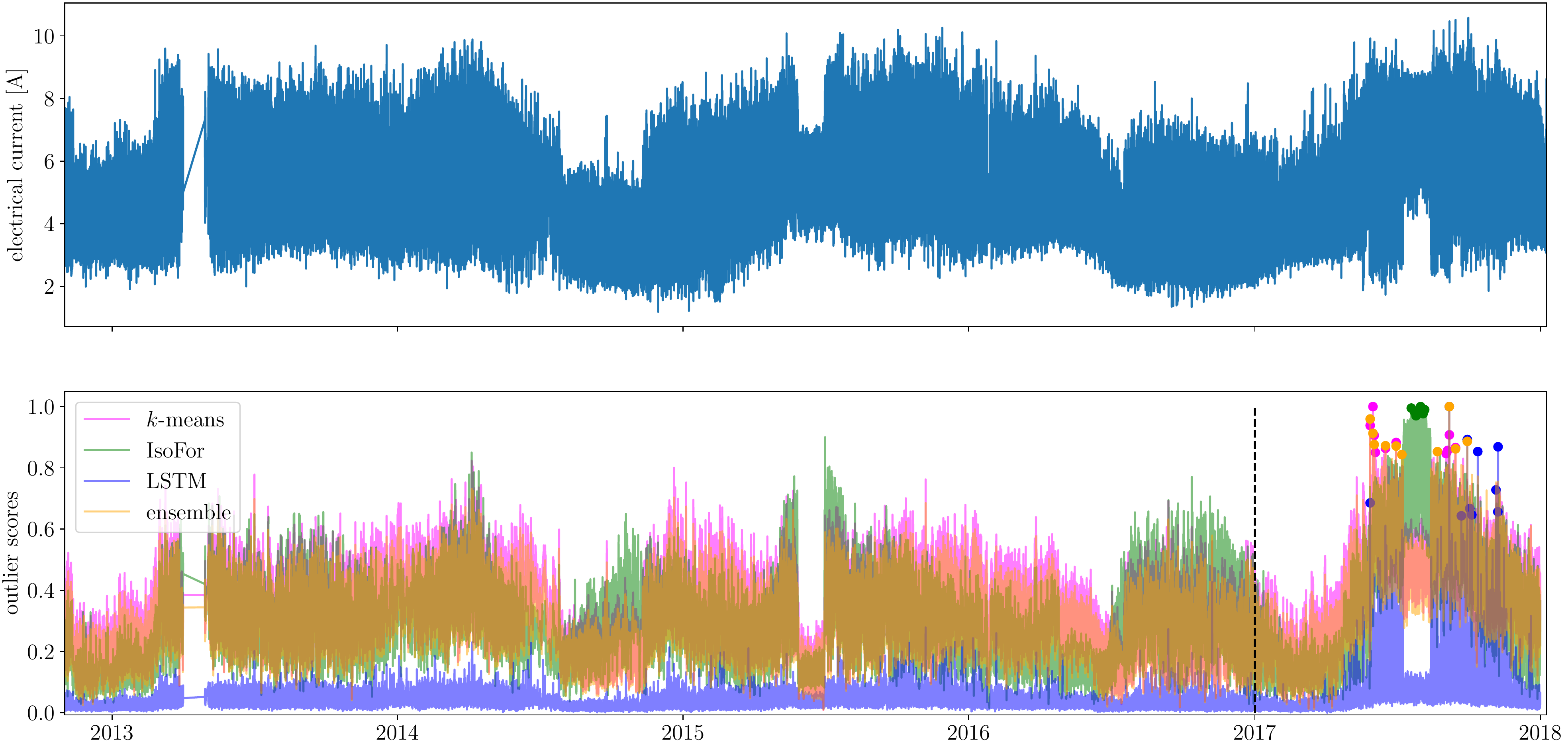}
    \caption{The most abnormal patterns in the data are detected in 2017.}
    \label{fig:2017}
\end{figure}

\begin{figure}
    \centering
    \includegraphics[width=\linewidth]{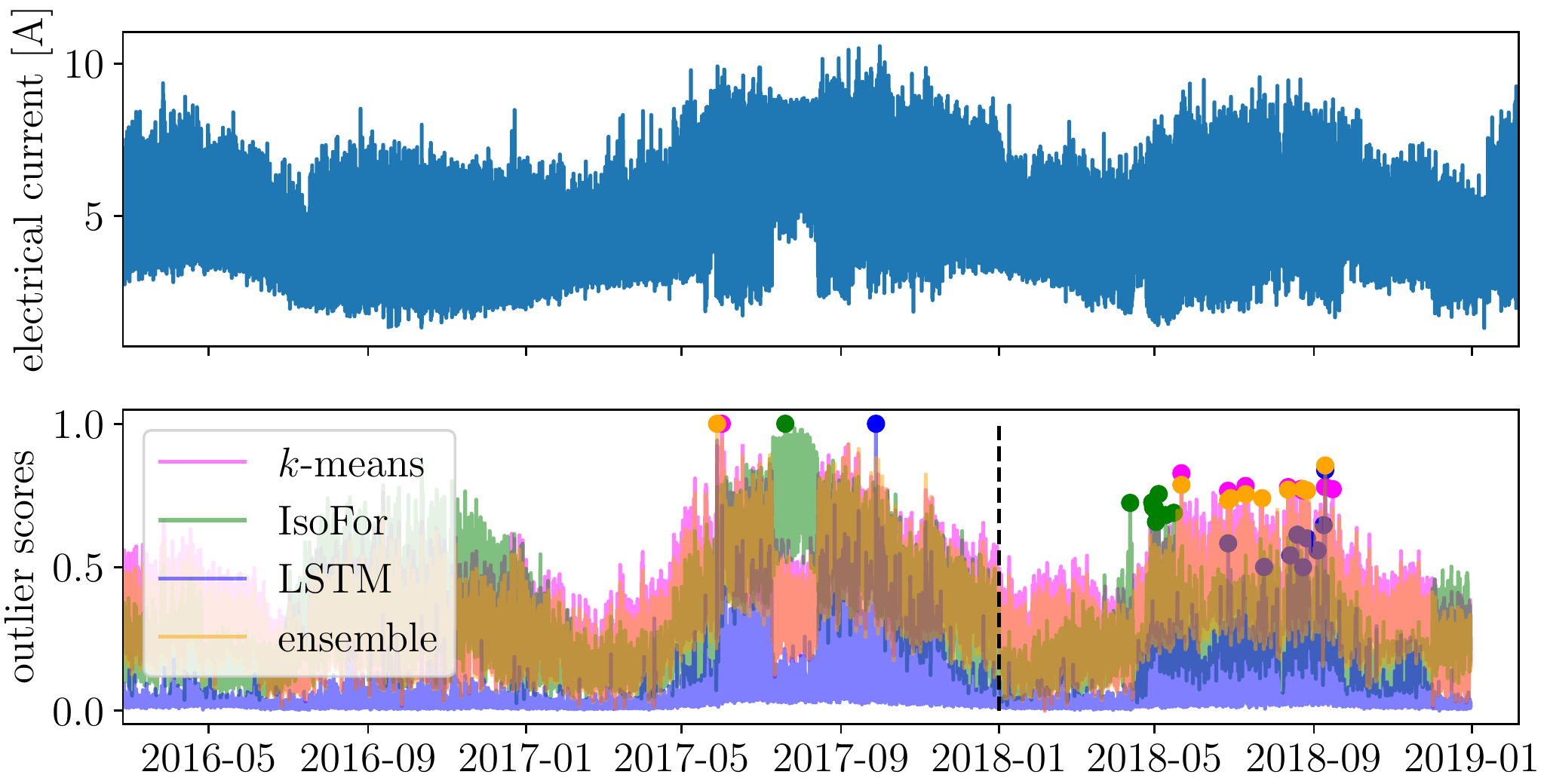}
    \caption{Outliers identified in the test set for the period $2018-2019$. The models, while successful, slightly struggle to adapt to new anomalous behavior.}
    \label{fig:2018}
\end{figure}

The most abnormal patterns in the data (including the highest peaks in MEX's operation) have been detected in the $2017-2018$ period. As shown in Fig.~\ref{fig:2017}, the conjunction in that period has been classified by the ensemble as normal, while some of the individual models (e.g., isolation forest) have reported outliers. These outliers, however, quite diverge from the expected behavior but still remain challenging to be discovered by the models. This is further evident in the last test period $2018-2019$ (Fig.~\ref{fig:2018}), where the anomalous records of 2017 are used in the model-training process. In this scenario, the models report higher error on the training data than on the testing set (i.e. larger outliers are identified in the training set), meaning that the models while successful, struggle a bit to swiftly adapt to new behavior. This is also in-line with our previous conjecture regarding the problem of potentially missing less obvious outliers, when large deviations are presented in the learning process. 

Another potential reason for this, is that the data in 2018 (and on wards) differs from the previous years. In 2018 MEX operations underwent a fundamental change, which required a software upload. This process involved performing no science operations for weeks, implementing the new methodology, as well as flying in never before used configurations. This had radical effects on the thermal power consumption. Which in turn makes learning from or comparing 2018 to any preceding year very difficult as it was so novel in so many ways. 

\section{Conclusion}\label{sec:conclusions}

In this paper we propose an end-to-end machine learning approach for outlier detection in telemetry data. More specifically, to ensure good performance, we construct a heterogeneous ensembles with constituents derived from three state-of-the-art unsupervised methods for outlier detection : $k$-means, isolation forest and long short-term memory (LSTM) autoencoders. We demonstrate the utility of the proposed approach on several tasks of identifying anomalous behavior in the electrical power consumption of the MEX's thermal subsystem. The results show that such an approach is able to accurately detect all major outliers (such as unusually high electrical currents, missing data and conjunctions). Moreover, this approach can provide additional insights into the spacecraft behavior during some rare events, such as Siding Spring comet avoidance maneuvers.

There are several potential directions for extending the work presented in this paper. First, we will focus on the problem of automatic threshold calibration, thus refining the conditions and circumstances for reporting an outlier. Second, we will investigate the task of contextual outlier detection, where the models for outlier detection are learned in the with respect to a given context. These contexts refer to an additional knowledge that can improve the predictions, such as scenarios when power-consumption fluctuations are expected during conjunction therefore small deviations won't be considered as outliers. Third, the methods presented in this paper are general, therefore can be extended to other subsystems of MEX and other spacecraft. Finally, we will extend our method to be able to detect also gradual changes (e.g., increasing trends), since until now, we were focusing only on the abrupt ones.

\section*{Acknowledgment}

The authors acknowledge the financial support of ESA through the project GalaxAI: Machine learning for space operations. Also, MP, TS, PP, NS and DK acknowledge the support of the Slovenian Research Agency through the research program No.~P2-0103 and research project No.~J2-9230. 



\bibliographystyle{IEEEtran}
\bibliography{references}
%

\end{document}